\documentstyle[12pt]{article}

\def\ci#1{\cite{#1}}
\def\bi#1{\bibitem{#1}}
\def\vak{V.A.\ Kosteleck\'y}

\def\ga{\gamma}
\def\de{\delta}
\def\ep{\epsilon}

\def\la{\lambda}

\def\si{\sigma}

\def\ps{\psi}
\def\om{\omega}

\def\De{\Delta}

\def\cE{{\cal E}}
\def\ea{{\it et al.}}
\def\fr#1#2{{{#1} \over {#2}}}

\def\ket#1{|{#1}\rangle}

\def\half{{\textstyle{1\over 2}}}
\def\frac#1#2{{\textstyle{{#1}\over {#2}}}}
\def\lsim{\mathrel{\rlap{\lower4pt\hbox{\hskip1pt$\sim$}}
    \raise1pt\hbox{$<$}}}
\def\gsim{\mathrel{\rlap{\lower4pt\hbox{\hskip1pt$\sim$}}
    \raise1pt\hbox{$>$}}}
\def\hydrogen{$H$}
\def\antihydrogen{$\overline{H}$}
\def\h{\hydrogen}
\def\ah{\antihydrogen}
\newcommand{\beq}{\begin{equation}}
\newcommand{\eeq}{\end{equation}}
\newcommand{\bea}{\begin{eqnarray}}
\newcommand{\eea}{\end{eqnarray}}
\newcommand{\rf}[1]{(\ref{#1})}
\def\ibid #1 #2 #3 {\it ibid., \rm {\bf #1}, #3 (19#2)}  %
\def\prd #1 #2 #3 {Phys.\ Rev.\ D {\bf #1}, #3 (19#2)}
\def\prl #1 #2 #3 {Phys.\ Rev.\ Lett.\ {\bf #1}, #3 (#2)}
\begin{document}
\setlength{\baselineskip}{5mm}

\noindent{\large\bf
Tests of Spacetime Symmetry with Particle Traps\footnote{Presented
at 23rd International Colloquium on
Group-Theoretical Methods in Physics, Dubna, \\
Russia, July-August 2000}
}\vspace{4mm}

\noindent{Neil Russell}\vspace{1mm}

\noindent{\small
Physics Department, Northern Michigan University,
Marquette, MI 49855, U.S.A.
}\vspace{4mm}

\abstract{Lorentz and CPT symmetry have been tested
at high precision in numerous experiments.
A general theoretical framework
incorporating possible Lorentz and CPT violation
in an extension of the standard model of particle physics
has been developed.
In this framework,
analyses of several precision experiments have been performed
to find unsuppressed symmetry-violating signals.
This paper discusses features of the theory,
presents results for trapped-particle systems,
and reports bounds from recent experiments.}

\section{Introduction}

Symmetry under the Lorentz and CPT transformations
is a property of the standard model of particle
physics\ci{sachs, cpt98}.
The possible violation of these symmetries
has been investigated in the context of an underlying theory
including also the gravitational interaction
\ci{kskp}.
Minuscule effects of Lorentz and CPT violation
might then be detectable in high-precision experiments.
The expected suppression of such effects would be
the ratio of a low-energy scale to the Planck scale.
These effects can be described by a
general standard-model extension
\ci{ck98}
that allows CPT and Lorentz violation
but retains the other conventional properties
of quantum field theory,
such as energy conservation, gauge invariance,
and renormalizability.

Sensitivity to certain effects in
the standard-model extension is known to exist in
a variety of experiments.
These include
tests with muons
\ci{BKLmuon},
experiments with kaons and other neutral mesons
\ci{mesons,ckpv},
studies of the baryon asymmetry
\ci{bckp},
measurements of cosmic birefringence
\ci{ck98,cfj,jk99},
clock-comparison experiments
\ci{cctests,kla,kljmp},
and investigations with spin-polarized solids
\ci{solids,heckel}.
This paper will review investigations
with low-energy trapped particles,
focusing on tests with Penning-traps
\ci{vd,gg99,hd99,rm99,bkr97,bkr98},
and tests involving the spectroscopy of
hydrogen and antihydrogen
\ci{bkr99}.

The extension of the SU(3)$\times$SU(2)$\times$U(1) standard model
and quantum electrodynamics
originates in the idea of spontaneous CPT and Lorentz breaking in an
underlying context such as string theory \ci{kskp}.
Violations of CPT and Lorentz symmetry are allowed in the
theory as couplings that can be experimentally bounded,
if not in fact measured.
In this context,
a particle with charge $q$ and mass $m$
is described by a four-component spinor field $\ps$
satisfying a Dirac equation with additional terms
\ci{ck98,bkr98}
\bea
\left( i \ga^\mu D_\mu - m - a_\mu \ga^\mu - b_\mu \ga_5 \ga^\mu -
\half H_{\mu \nu} \si^{\mu \nu} + i c_{\mu \nu} \ga^\mu D^\nu +
i d_{\mu \nu} \ga_5 \ga^\mu D^\nu \right.
&& \nonumber \\ \left. + i e_{\mu}D^\mu - f_\mu \ga_5 D^\mu +
\half i g_{\mu\nu\la}\si^{\mu\nu} D^\la \right) \ps = 0 && .
\label{dirac}
\eea
In this equation,
$A^\mu$ is the electromagnetic potential and
$i D_\mu\equiv i \partial_\mu - q A_\mu$.
The symmetry violations are parametrized by
a set of effective coupling constants
$a_\mu$, $b_\mu$,
$c_{\mu \nu}$, $d_{\mu \nu}$,
$e_\mu$, $f_\mu$, $g_{\mu\nu\la}$,
and $H_{\mu \nu}$;
$c_{\mu\nu}$ and $d_{\mu\nu}$ are traceless,
$g_{\mu\nu\la}$ is antisymmetric in the first two indices,
$H_{\mu\nu}$ is antisymmetric,
and all are real.
The $a_\mu$, $b_\mu$, $e_\mu$, $f_\mu$, and $g_{\mu\nu\la}$ terms
break CPT,
while those involving
$H_{\mu\nu}$, $c_{\mu\nu}$, and $d_{\mu\nu}$ preserve it.
All of them are observer Lorentz covariant, but break
particle Lorentz symmetry.
Using a suitable field redefinition
it is possible at first order to eliminate all the $e_\mu$ and $f_\mu$
terms and some of the $g_{\mu\nu\la}$,
so we set $e_\mu$, $f_\mu$, and $g_{\mu\nu\la}$ equal to
zero without any significant loss of generality
\ci{ck98}.

\section{Symmetry Tests with Penning Traps}
The Penning trap is a device that uses a uniform magnetic field
and a quadrupole electric field to confine charged particles.
The quantum behavior of the trapped particles can be studied with
a high degree of precision and control.
For example,
it is possible to capture
a single electron, positron, proton or antiproton
and measure its motional frequencies
over a period of several months.
Two of these oscillation frequencies, which can be measured
with precision better than a part in $10^8$,
are the cyclotron frequency $\om_c$ and the anomaly frequency $\om_a$.
In the context of the standard-model extension,
violations of Lorentz and CPT symmetry
yield shifts of these frequencies.
For an electron or positron,
the leading-order shifts are
\bea
\om_c^{e^-} &\approx& \om_c^{e^+}
\approx (1 - c_{00}^e - c_{11}^e - c_{22}^e) \om_c
\quad , \label{wcelec} \\
\om_a^{e^\mp} &\approx&
\om_a \mp 2 b_3^e + 2 d_{30}^e m_e + 2 H_{12}^e
\quad . \label{waelec}
\eea
In this notation,
the superscript $e^{\pm}$ refers to the positron or electron,
and $\om_c^{e^\mp}$, $\om_a^{e^\mp}$
represent the shifted frequencies.
For other particles,
for example protons and antiprotons,
the expressions have appropriately modified superscripts.

\subsection{Frequency-comparison tests}
A category of Lorentz and CPT tests
involves the comparison of frequencies that
are equal in the conventional standard model of particle physics.
Included among these are the cyclotron and anomaly frequencies of
particles in Penning traps
as compared with the corresponding antiparticle frequencies.
In the standard-model extension,
the electron-positron differences for the
cyclotron and anomaly frequencies
can be found from Eqs.~\rf{wcelec} and \rf{waelec}:
\beq
\De \om_c^e \equiv \om_c^{e^-} - \om_c^{e^+} \approx 0 \quad , \qquad
\De \om_a^e \equiv \om_a^{e^-} - \om_a^{e^+} \approx - 4 b_3^e \quad .
\label{delwce}
\eeq
It follows that the dominant signal
for CPT violation in Penning-trap experiments
is a difference between the electron and positron anomaly frequencies.
The $b_3$ coupling violates both Lorentz and CPT symmetry,
so no leading-order contributions appear
from CPT-preserving but Lorentz-breaking terms.
Leading-order signals
in cyclotron-frequency comparisons
are suppressed in this context.
A figure of merit for the test
can be introduced as the ratio of
a CPT-violating electron-positron energy-level difference
and the basic energy scale
\ci{bkr97}
\beq
r^e_{\om_a} \equiv \fr{|{\cal E}_{n,s}^{e^-} - {\cal E}_{n,-s}^{e^+}|}
                        {{\cal E}_{n,s}^{e^-} }
\quad . \label{re}
\eeq
In this expression,
$\cE^{e^-}_{n,s}$ and $\cE^{e^+}_{n,s}$ are
energy eigenvalues of the full Penning-trap hamiltonians,
with principal quantum numbers $n=0,1,2,\ldots$
and spin $s=\pm 1$.
For the nonrelativistic regime of relevance here,
$\cE^{e^-}_{n,s}$ is essentially the rest mass $m_e$
and consequently Eq.~\rf{re} reduces to
\beq
r^e_{\om_a} \approx \fr{| \De \om_a^e |}{2 m_e}
\approx \fr{|2 b_3^e |}{m_e}
\quad . \label{reb}
\eeq
One may estimate, for example,
that if the anomaly frequencies were measured
to an absolute precision of about 2~Hz,
then a bound
$r^e_{\om_a} \lsim 10^{-20}$ would be placed.

The Penning-trap group of Hans Dehmelt
at the University of Washington in Seattle
recently published a result
based on this type of anomaly-frequency comparison
\ci{hd99}.
A bound of
\beq
r^e_{\om_a} < 1.2 \times 10^{-21}
\eeq
was found
from a reanalysis of earlier data for
$g-2$ experiments comparing
single trapped electrons and single trapped positrons.

Previous CPT tests done with the Penning trap include
comparisons of the gyromagnetic ratios of electrons and positrons.
For example,
one of the conventional figures of merit for CPT symmetry is
\ci{vd}
\beq
\left|\fr{(g_- - g_+)}{g_{\rm av}}\right|
\lsim 2 \times 10^{-12}
\quad . \label{rg}
\eeq
However,
in the framework of the standard-model extension,
CPT is broken without
affecting the electron or positron gyromagnetic ratios.
Thus,
the theoretical value of the figure of merit in Eq.~\rf{rg}
would be zero even if CPT were broken,
and this figure of merit is
unsuitable in this theoretical context.

While it might appear from Eq.~\rf{delwce}
that comparisons of cyclotron frequencies
are insensitive to the CPT and Lorentz violations
in the standard-model extension,
this is in fact not entirely true.
An experiment
\ci{gg99}
by the group of Gerald Gabrielse at Harvard University
compared the cyclotron frequencies of antiprotons and hydrogen ions
and obtained a bound on a combination of Lorentz-violation couplings.
This choice of ions was made to eliminate the difficulties
of precisely reversing the electrode potentials when particles of
opposite charge are loaded into the trap
\ci{gg95}.
Since the hydrogen ion $H^-$ and the antiproton
both have negative charges,
no electric-field reversal is necessary,
and both particles can be simultaneously trapped.
Established precision measurements
of the electron mass and the $H^-$ binding energy
can be used to estimate
the theoretical value of the difference
$\De \om_c^{H^-} \equiv\om_c^{H^-} - \om_c^{\bar p}$
in conventional quantum theory.
With these corrections for the two electrons in
the $H^-$ ion,
the experiment allows a comparison of
the proton component of the $H^-$ ion
with the antiproton.

In the context of the standard-model extension,
this comparison of cyclotron frequencies
is shifted at leading order
by a combination of Lorentz-violating couplings
\ci{bkr98}.
A model-independent figure of merit
\beq
r^{H^-}_{\om_c} \lsim \left| \fr{\De\om_{c,{\rm th}}^{H^-}}{m_p}\right|
\eeq
can be defined.
One of the results of the Gabrielse experiment
was the bound
\beq
r^{H^-}_{\om_c} \lsim 4 \times 10^{-26}
\quad .
\label{Hminbound}
\eeq
Within the standard-model extension,
this result limits a combination
of Lorentz-violating, CPT-preserving couplings,
including $c_{00}^e$ and $c_{00}^p$
which are not accessible in other similar experiments.

\subsection{Sidereal-variation tests}
The CPT- and Lorentz-violating couplings in
the standard-model extension
are constant vacuum expectation values of
tensorial objects in a more fundamental theory.
The physics of these couplings is approximately analogous
to that of electrodynamics in macroscopic media
\ci{ck98}.
Earthbound experiments sensitive to
these minuscule couplings
could seek to detect
oscillations in experimental observables
due to the rotation of the earth.
These would be expected at various multiples
of the earth's sidereal frequency.

The conventional standard model
predicts that
the measured Penning-trap frequencies for an electron
should remain constant
provided the magnetic and electric fields remain constant.
In the context of the standard-model extension,
variations in the electron frequencies
can be found from Eqs.~\rf{wcelec} and \rf{waelec}
by noting that the indices in these expressions
are given in the laboratory coordinate system
that rotates against the fixed stars.
A more complete discussion of tests of this type
with the Penning trap is discussed in
Ref.~\ci{bkr98}.

For a single electron in a Penning trap,
the anomaly frequency $\om_a^e$
is expected to have a variation
with frequency equal to one sidereal day
due to the index structure in Eq.~\rf{waelec}.
These indices are defined in terms of the magnetic field direction,
which is fixed in the laboratory,
but which rotates in the celestial equatorial coordinate system
\ci{kla}.

A model-independent figure of merit
sensitive to the present effects
may be defined in terms of the quantity
\ci{dubin}
\beq
\De^{e}_{\om_a^{e^-}} \equiv
\fr {|{\cE}_{0,+1}^{e^-} - {\cE}_{1,- 1}^{e^-}|}
{{\cE}_{0,-1}^{e^-}}
\quad , \label{Deleorpdnlom}
\eeq
which is essentially the ratio of
the anomaly frequency to the rest mass of the electron.
The amplitude of sidereal variations in this dimensionless quantity
defines a figure of merit
$r^e_{\om_{a}^-,\rm sidereal}$
for this type of Lorentz-violating effect.

Data from an experiment confining a single electron in a Penning trap
for several weeks
have recently been reanalyzed by Mittleman of
the Dehmelt trapping group.
To search for sidereal variations,
the data were partitioned into sidereal bins
determined by the direction of the magnetic field.
The bound obtained~\ci{rm99} is
\beq
r^e_{\om_a^-,{\rm sidereal}} \leq 1.6 \times 10^{-21} \quad .
\eeq
In the present context,
this constrains a combination of Lorentz-violating couplings,
some of which also violate CPT.

\section{Hydrogen and Antihydrogen}
The hydrogen atom is
one of the most studied systems in physics.
Comparison of hydrogen (\h) with antihydrogen (\ah)
requires the availability of antihydrogen atoms
in quantities suitable for
precision spectroscopy.
As of October 2000,
this is not a reality,
although about a dozen events consistent with \ah\ creation
were reported in a 1995 experiment at CERN
\ci{oelert}
and another dozen in a 1996 Fermilab experiment
\ci{mandel}.
Current efforts by two experimental groups
\ci{gab2,holz}
using the antiproton decelerator at CERN are underway to
improve on these initial experiments
and eventually create trapped antihydrogen for precision studies.
Confinement would be within magnetic traps
like the Ioffe-Pritchard trap
\ci{i-p}.

An analysis of the spectra of
\hydrogen\ and \antihydrogen\
in the context of the standard-model extension
has been done
for both free and trapped atoms
\ci{bkr99}.

\subsection{Comparisons of hydrogen and antihydrogen}
One of the spectral lines of importance
is the two-photon 1S-2S transition
because of its eventual expected measurement precision
of a part in $10^{18}$.
So far,
relative precisions for this transition
stand at a few parts in $10^{14}$
\ci{hansch} for free hydrogen
and a few parts in $10^{12}$
\ci{cesar} for trapped hydrogen.
The possible signals affecting the 1S-2S transition in free hydrogen
in the context of the standard-model extension
have been found to be suppressed by
at least two powers of the fine-structure constant
\ci{bkr99}.

Turning to the analysis of trapped hydrogen and antihydrogen,
it is useful to consider the case where the trap has a
magnetic bias field $B$
that splits the 1S and 2S levels into four hyperfine Zeeman levels,
denoted in order of increasing energy by $\ket{a}_n$, $\ket{b}_n$,
$\ket{c}_n$, $\ket{d}_n$,
with principal quantum number $n=1$ or $2$, for both
\hydrogen\ and \ah.
Only transitions involving the
$\ket c$ and $\ket d$ are relevant
because these are the two trapped states.
For small values of the $B$ field,
transitions between the $\ket{d}_1$ and $\ket{d}_2$ states
are field independent.
So, by comparing the frequency $\nu^H_d$ for the 1S-2S transition
$\ket{d}_1 \rightarrow \ket{d}_2$ in \hydrogen\
with the corresponding frequency
$\nu^{\overline{H}}_d$ in \ah,
effects due to magnetic-field instability and
inhomogeneity would be minimized.
However,
the analysis again shows
$\de \nu^H_d = \de \nu^{\overline{H}}_d \simeq 0$ at leading order.
There are no unsuppressed frequency shifts in this \hydrogen\ transition
or the corresponding \ah\ transition.

An alternative would be consideration of the 1S-2S transition between the
states $\ket{c}_1$ and $\ket{c}_2$ in \hydrogen\ and \ah.
In the present theoretical context,
an unsuppressed frequency shift does indeed occur
in this transition
because the $n$ dependence in the hyperfine splitting
produces a spin-mixing difference
between the 1S and 2S levels.
The leading-order frequency shift is field dependent
with a maximum at about $B \simeq 0.01$ T.
However,
the strong field gradient at this value of $B$
could severely limit the precision.

Another possibility for investigating
CPT and Lorentz-violation in the context of the standard-model extension
is to consider hyperfine transitions in the 1S ground state of \h.
The analysis is done by considering the perturbative shifts in the
energy levels of the relativistic \h\ atom using the
Dirac equation \rf{dirac}.
The CPT- and Lorentz-violating couplings
give rise to field-dependent energy shifts
of the $\ket a$ and $\ket c$ hyperfine levels
and field-independent shifts of
the $\ket b$ and $\ket d$ hyperfine levels in
the 1S ground state of \h.

An interesting case is the
$\ket{d}_1 \longrightarrow \ket{c}_1$ transition,
also known as the $F$=1, $\De m_F= \pm 1$ transition.
While this transition can be measured at various frequencies,
there is some advantage
from a theoretical standpoint
of selecting a magnetic field of about 0.65~T,
since this minimizes suppression effects.
At this field value,
the leading-order difference in the frequencies
$\nu_{c \rightarrow d}^H$ and
$\nu_{c \rightarrow d}^{\overline{H}}$
is
$\De \nu_{c \rightarrow d} \equiv
\nu_{c \rightarrow d}^H - \nu_{c \rightarrow d}^{\overline{H}}
\approx - 2 b_3^p / \pi$.
Within the context of the standard-model extension,
this \h-\ah\ comparison
isolates the CPT-violating coupling
$b_3^p$ for the proton and is therefore of interest
as a clean CPT test.
An appropriate model-independent figure of merit
for any experimental comparison
 of this frequency in \h\ and \ah\
can be defined by
\ci{bkr99}
\bea
r^H_{rf,c \rightarrow d} & \equiv &
{|({\cal E}_{1,d}^H - {\cal E}_{1,c}^H)
- ({\cal E}_{1,d}^{\overline{H}} - {\cal E}_{1,c}^{\overline{H}})|}/
{{\cal E}_{1,{\rm av}}^H}
\nonumber \\
&\approx &
2\pi |\De \nu_{c \rightarrow d}| /m_H
\quad ,
\label{rrf}
\eea
where the $\cE$ denote relativistic energies
for hydrogen and antihydrogen in the ground-state hypefine level
and
where $m_H$ is the atomic mass of \h.
Assuming  a frequency resolution of about 1 mHz
would be possible with both species of particles,
an upper bound of
$r^H_{rf,c \rightarrow d} \lsim 5 \times 10^{-27}$
can be estimated.
The bound on the CPT- and Lorentz-violating coupling $b_3^p$
would be
$|b_3^p| \lsim 10^{-18}$ eV,
an improvement of four orders of magnitude
over bounds estimated for 1S-2S transitions.

Direct comparisons of frequencies in hydrogen with corresponding
frequencies in antihydrogen
are of course not possible until \ah\ is readily available for
spectroscopic experiments.

\subsection{Sidereal-variation tests in Hydrogen}

The properties of the couplings in the standard-model extension
mean that frequencies such as the ground-state hyperfine transition
$\De \nu_{c \rightarrow d}$ in \h\
should have small variations due to the sidereal rotation of the earth.
Thus interesting bounds can be placed
on certain combinations of couplings
using only hydrogen.

Such an experiment,
searching for sidereal variations in
the $F=1$, $\De m_F=\pm 1$ transition
of a \h\ maser has recently been completed at
the Harvard-Smithsonian Center for Astrophysics
\ci{8230}.
The maser was run with a weak bias magnetic field of 0.6 mG,
for which the corresponding Zeeman frequency is about 850 Hz.
A double resonance technique was used to monitor this frequency,
with a resolution of about $0.37$ mHz.
This bounds sidereal variations
at the level of $1.5\times 10^{-27}$ GeV.
In the context of the standard-model extension,
the parameters bounded here are a combination
of electron and proton parameters,
\beq
\left |\tilde{b}_3^p + \tilde{b}_3^e \right | \leq 2 \pi \de\nu_Z
\quad ,
\eeq
where
$\tilde{b}_J = b_J - m_e d_{0J} - \half \ep_{JKL}H_{KL}$
for both superscripts,
and
$\de \nu_Z$ is the sidereal-frequency modulation
of the Zeeman frequency
\ci{kla}.

A bound of $10^{-29}$ GeV has independently been placed on
the electron parameter
$\tilde{b}_J^e$
using a spin-polarized torsion pendulum
\ci{heckel}.
Here, $J$ refers to spatial components
in the nonrotating celestial coordinate system.
This result was obtained from a reanalysis of data from the E\"ot-Wash II
experiment conducted at the University of Washington.
Combining the hydrogen maser result mentioned above
and this tight bound on the electron parameter,
it can be inferred that
the hydrogen maser experiment places the bound
\beq
\tilde{b}_J^p \leq 10^{-27} {\rm GeV} \quad .
\eeq

\section{Related tests}
Clock-comparison tests
have been used to study Lorentz symmetry
and have resolutions
of less than a $\mu$Hz,
several orders of magnitude better than for
the hydrogen-maser system.
However,
analysis of effects within the framework of the standard model extension
is far more complex than for hydrogen,
and has to rely on various nuclear models
\ci{kla}.
In comparison,
the parameter combination
$\tilde{b}_J^p$
bounded in the hydrogen-maser experiment
is considerably cleaner than other comparable bounds from
clock-comparison experiments,
such as for the $^{199}$Hg/$^{133}$Cs system
\ci{kla,berglund}.

An experiment with a dual-species $^{129}$Xe$/^3$He maser
has recently placed a limit on a combination of CPT- and Lorentz-violating
parameters within the standard-model extension
\ci{7049}.
With a resolution of about 45 nHz,
the bound is
\beq
\tilde{b}_\perp^n \equiv \sqrt{(\tilde{b}_X^n )^2 + (\tilde{b}_Y^n )^2}
=(4.0\pm 3.3)\times10^{-31} {\rm GeV} \quad ,
\eeq
consistent with no Lorentz- and CPT-violating effects under
reasonable statistical assumptions.
This result, obtained in Walsworth's laboratory
at the Harvard-Smithsonian Center
for Astrophysics, improves on the tightest previous limits
\ci{kla}
for the CPT- and Lorentz-violating couplings of the neutron
by a factor of more than six.
Indications are that an improvement of about an order of magnitude
will be possible with further refinements.
In addition, a new experiment under development using
a two-species $^{21}$Ne/$^3$He maser
\ci{stoner}
is expected to improve the resolution by a further order of magnitude.

Also of interest are recent bounds on Lorentz symmetry
from neutrino-oscillation investigations
\ci{lisi}.

\section*{Acknowledgments}
I thank the organizers of the
International Colloquium on Group-Theoretical Methods in Physics
for their invitation to this enjoyable conference.
Northern Michigan University contributed towards
travel expenses.
This work was partially supported by
the United States Department of Energy
under grant DE-FG02-91ER40661.

\setlength{\parindent}{0mm}

\end{document}